\documentclass{article}
\usepackage{amssymb}

\def\rset{\mathbb{R}}
\def\zset{\mathbb{Z}}

\def\ds{\displaystyle}

\def\pd#1#2{{\ds\partial #1\over\ds\partial #2}}

\newtheorem{thm}{Theorem}
\newtheorem{lm}[thm]{Lemma}
\newtheorem{prop}[thm]{Proposition}

\title{On Periodic Solutions of a Hamilton--Jacobi Equation with
Periodic Forcing\footnote{See also: \textit{Mat.\ Sbornik} \textbf{190}:10
(1999) 87--104 (in Russian). English translation: \textit{Sbornik:
Mathematics}, \textbf{190}:10 (1999) 1487--1504.}}

\author{A.N.~Sobolevski\u\i}
\date{Moscow State University\\
e-mail: \texttt{ansobol@idempan.phys.msu.su}}

\begin{document}

\maketitle

\begin{abstract}
We prove that a Hamilton--Jacobi equation in 1D with periodic forcing
has a set of generalized solutions such that each solution
is a sum of linear and continuous periodic functions; we also give a
condition of uniqueness of such solution in terms of Aubry--Mather theory.
\end{abstract}

\section{Introduction}

Consider a Hamilton--Jacobi equation
\begin{equation}
   \pd{S}{t} + H_0\left(\pd{S}{x}\right) + U(t,x) = 0,
\label{HJE}
\end{equation}
where $x \in \rset$ and the functions $H_0(p)$, $p \in \rset$, and
$U(t,x)$ satisfy the following assumptions:
\begin{itemize}
\item[$(H_1)$] $H_0(p)$ is of class $C^1$ and its derivative $H'_0(p)$ is
strictly growing: $H'_0(p_2) > H'_0(p_1)$ iff $p_2 > p_1$, and locally
Lipschitzian: for any $P > 0$ there exists $C > 0$ such that for all $p_1$,
$p_2 \in [-P, P]$ $|H'_0(p_1) - H'_0(p_2)| \le C |p_1 - p_2|$.
\item[$(H_2)$] For any $N > 0$ there exists $P > 0$ such that if $p \in
\rset$, $|p| > P$, then $H_0(p) > N|p|$.
\item[$(H_3)$] $U(t,x)$ is of class $C^1$ and its derivative $\partial U/
\partial x$ is Lipschitzian.
\item[$(H_4)$] $U(t,x)$ is periodic with respect to $t$ and $x$: $U(t+1,x) =
U(t,x+1) = U(t,x)$ for all $(t,x) \in \rset^2$.
\end{itemize}
Here and below, $({}')$ denotes an ordinary derivative and `periodic' means
`periodic with period 1 in each variable'; we shall refer to the whole set
of~$(H_1)$--$(H_4)$ as assumptions~$(H)$.

Suppose $U(t,x) = 0$; then equation~(\ref{HJE}) has a one-parameter family
of classical solutions $S^a(t,x) = ax - H_0(a)t$. In this paper we prove
the following
\medskip
\par\noindent\textbf{Theorem I}
\emph{If $H_0(p)$ and $U(t,x)$ satisfy assumptions $(H)$, then for any $a
\in \rset$ there exists a continuous periodic function $s^a(t,x)$ such that
the function}
\begin{equation}
   S^a(t,x) = ax - H(a)t + s^a(t,x)
\label{defSa}
\end{equation}
\emph{is a viscosity solution of equation~(\ref{HJE}) (for definition of a
viscosity solution to a Hamilton--Jacobi equation, see~\cite{CL}). Here
$H(a)$ is a convex function satisfying}
\begin{equation}
   \min_{(t,x) \in \rset^2} U(t,x)
   \le H(a) - H_0(a)
   \le \max_{(t,x) \in \rset^2} U(t,x).
\label{estHH0}
\end{equation}
\medskip
\par\noindent
In section~2 we introduce an explicit representation of a viscosity
solution to Cauchy problem for equation~(\ref{HJE}) by the well-known
Lax--Ole\u{\i}nik formula. In the sequel, this formula is used as
a substitute for the general definition of a viscosity solution.

Let $a_0 \in \rset$ and $S^{a_0}(t,x)$ be some viscosity solution
to~(\ref{HJE}) of the form~(\ref{defSa}) with $a = a_0$. For any $c \in
\rset$, $S^{a_0}(t,x) + c$ is another viscosity solution to~(\ref{HJE}) of
the form~(\ref{defSa}). The following statement specifies the case when the
converse is true, i.e., when any viscosity solution of the
form~(\ref{defSa}) with $a = a_0$ differs from $S^{a_0}(t,x)$ by a
constant:
\medskip
\par\noindent\textbf{Theorem II}
\emph{Let $s^a(t,x)$ be the continuous periodic function of theorem~I and
$\omega(a)$ be the corresponding rotation number (see section~6). If
$\omega(a)$ is irrational, then $s^a(t,x)$ is unique up to an additive
constant.}
\medskip

This paper is organized as follows. In sections~2 and~3 we prove some
auxiliary results. Section~4 is devoted to the proof of theorem~I. In
sections~5 and~6 we introduce a many-valued map associated with a
solution~(\ref{defSa}) and study the connection between properties of such
solutions and Aubry--Mather theory \cite{A,AlD,M}. Section~7 concludes the
proof of theorem~II.

The results of this paper were announced in \cite{S}; they extend and
strengthen some results of~\cite{JKM}.

Similar results were obtained by Weinan~E \cite{E} independently of this
author.

The author is grateful to Prof.~Ya.G.~Sinai for setting of the problem
and constant attention to this work and to Prof.~Weinan E for the
opportunity to read a preprint of his paper.

\section{The value function of the action functional}

\emph{A Legendre transform} of the function $H_0(p)$ is a $C^1$ function
$L_0(v)$ such that $L'_0(H'_0(p)) = p$. If $H_0(p)$ satisfies
assumptions~$(H_1)$ and~$(H_2)$, then $L_0(v)$ is finite for all $v \in
\rset$, has a continuous strictly growing derivative, and satisfies the
condition obtained from~$(H_2)$ by substitution of $L_0$ for $H_0$.
\emph{The Lagrangian} associated with
equation~(\ref{HJE}) is a function $L(t,x,v)$ of the form
\begin{equation}
   L(t,x,v) = L_0(v) - U(t,x).
\label{defL}
\end{equation}

Let $x$, $y$, $s$, $t \in \rset$, $s < t$. The set of all absolutely
continuous functions $\xi$:~$[s,t] \to \rset$ such that $\xi(s) = y$,
$\xi(t) = x$, and $L(\cdot, \xi(\cdot), \xi'(\cdot))$ is Lebesgue
integrable is called the set of \emph{admissible trajectories} (or
\emph{trajectories} for short) and is denoted by $\Omega(s,y; t,x)$. The
functional defined on $\Omega(s,y; t,x)$ by the formula
\begin{equation}
   \mathcal{L}(s,y;t,x)[\xi]
   = \int_s^t L(\tau, \xi(\tau), \xi'(\tau))\, d\tau
   = \int_s^t [L_0(\xi'(\tau)) - U(\tau,\xi(\tau))]\,d\tau
\label{delcalL}
\end{equation}
is called \emph{the action functional} associated with the Lagrangian
$L(t,x,v)$. It follows from assumptions~$(H)$ that this functional is
bounded from below on the set $\Omega(s,y;t,x)$. The infimum of
$\mathcal{L}(s,y;t,x)$ over $\Omega(s,y;t,x)$ is called the \emph{value
function} of the action functional and is denoted by $L(s,y;t,x)$.

\begin{lm}[see~{\cite[sections~2.6, 9.2, and 9.3]{C}}]
Suppose $H_0(p)$ and $U(t,x)$ satisfy assumptions~$(H)$; then there exists
a trajectory $\xi_0 \in \Omega(s,y;t,x)$ such that
$\mathcal{L}(s,y;t,x)[\xi_0] = L(s,y;t,x)$, $\xi_0$ is of class $C^1$ in
$[s,t]$, and there exists a $C^1$ function $\lambda$:~$[s,t] \to \rset$
such that
\begin{equation}
   \lambda(\tau) = L'_0(\xi'_0(\tau)),\qquad
   \lambda'(\tau) = - \pd{U(\tau,\xi_0(\tau))}{\xi_0},\qquad
   s < \tau < t.
\label{Euler}
\end{equation}
For all $R$, $s$, $t \in \rset$ such that $R > 0$, $s < t$ and all $x$, $y
\in \rset$ such that $|x - y| \le R(t - s)$ there exists a positive
constant $C_1 = C_1(R,s,t)$ such that $|\xi'_0(\tau)| \le C_1$, $s \le \tau
\le t$ for any $\xi_0 \in \Omega(s,y;t,x)$ such that
$\mathcal{L}(s,y;t,x)[\xi_0] = L(s,y;t,x)$.
\label{l:Cesari}
\end{lm}
The trajectory $\xi_0$ is called \emph{a minimizer} of the functional
$\mathcal{L}(s,y;t,x)$.

Let $a \in \rset$ and $s_0(x)$ be a continuous periodic function. The
viscosity solution of a Cauchy problem for equation~(\ref{HJE}) with the
initial data
\begin{equation}
   S_0(x) = ax + s_0(x)
\label{HJEini}
\end{equation}
is given by \emph{the Lax--Ole\u{\i}nik formula} (see,
e.g.,~\cite[section~11.1]{L82})
\begin{equation}
   S(t,x) = \inf_{y \in \rset} (S_0(y) + L(0,y;t,x)).
\label{FLO}
\end{equation}

\begin{lm}
The function $L(s,y;t,x)$ is diagonally periodic:
\begin{equation}
   L(s,y+1;t,x+1) = L(s+1,y;t+1,x) = L(s,y;t,x).
\label{diagper}
\end{equation}
\label{l:diagper}
\end{lm}
This lemma follows from periodicity of $U(t,x)$ in $t$ and $x$.
\medskip
\par\noindent\textbf{Corollary}
\emph{Let $S(t,x)$ be the solution of the Cauchy problem for
equation~(\ref{HJE}) with the initial data~(\ref{HJEini}); then $S(t,x) =
ax + s(t,x)$, where $s(t,x)$ is periodic in $x$.}
\medskip
\par\noindent
\textsc{Proof.}
$$
\begin{array}{rl}
   s(t,x+1)
   &\ds\!\!\!\!= S(t,x+1) - a(x+1) = \\
   &\ds\!\!\!\!= \inf_{y \in \rset}\,(s_0(y)+a(y-(x+1))+L(0,y;t,x+1)) = \\
   &\ds\!\!\!\!= \inf_{y \in \rset}\,(s_0(y-1)+a((y-1)-x))+L(0,y-1;t,x)) = \\
   &\ds\!\!\!\!= S(t,x) - ax = s(t,x).
\end{array}
$$
\hfill\hfill$\square$

\begin{lm}
For all $r$, $s$, $t$, $x$, $y \in \rset$, $s < r < t$,
\begin{equation}
   L(s,y;t,x) = \min_{z \in \rset}\,(L(s,y;r,z) + L(r,z;t,x)).
\label{Bellman}
\end{equation}
\label{l:Bellman}
\end{lm}
This is a variant of the well-known Bellman's principle of optimality.

\begin{lm}
The function $L(s,y;t,x)$ is everywhere finite and satisfies inequalities
\begin{equation}
   -M \le {1\over t-s}L(s,y;t,x) - L_0\left({x-y\over t-s}\right) \le -m,
\label{updown}
\end{equation}
where $m = \min_{(t,x) \in \rset^2}\,U(t,x)$, $M = \max_{(t,x) \in
\rset^2}\,U(t,x)$.
\label{l:updown}
\end{lm}
\textsc{Proof.} Consider the functional
$$
   \widehat{\mathcal{L}}(s,y;t,x)[\xi]
   = \int_s^t L_0(\xi'(\tau))\,d\tau.
$$
Recall that $L_0(v)$ is a $C^1$ convex function with strictly icreasing
derivative. Then it can be shown in the usual way that the trajectory
$\widehat\xi(\tau) = x(\tau-s)/\mbox{$(t-s)$} 
+ y(t-\tau)/(t-s)$, $s \le \tau \le t$, is a minimizer of this functional
in the set $\Omega(s,y;t,x)$ and
$$
   \widehat{\mathcal{L}}(s,y;t,x)[\widehat\xi]
   = (t - s) L_0\left({x - y\over t - s}\right).
$$

Let $\xi_0$ be a minimizer of the functional $\mathcal{L}(s,y;t,x)$ in
the class $\Omega(s,y;t,x)$, i.e., $\mathcal{L}(s,y;t,x)[\xi_0] =
L(s,y;t,x)$. We have
$$
\begin{array}{l}
   \ds L(s,y;t,x) - (t-s)L_0\left({x-y\over t-s}\right)
   = \mathcal{L}(s,y;t,x)[\xi_0]
   - \widehat{\mathcal{L}}(s,y;t,x)[\widehat\xi]) = \\
   \ds\qquad= -\int_s^t U(\tau,\xi_0(\tau))\,d\tau
   + (\widehat{\mathcal{L}}(s,y;t,x)[\xi_0]
   - \widehat{\mathcal{L}}(s,y;t,x)[\widehat\xi]) \ge \\
   \ds\qquad\ge (t-s)\left(-\max_{(t,x) \in \rset^2} U(t,x)\right)
\end{array}
$$
and
$$
\begin{array}{l}
   \ds L(s,y;t,x) - (t-s)L_0\left({x-y\over t-s}\right)
   = \mathcal{L}(s,y;t,x)[\xi_0]
   - \widehat{\mathcal{L}}(s,y;t,x)[\widehat\xi] \le \\
   \ds \qquad\le \mathcal{L}(s,y;t,x)[\widehat\xi]
   - \widehat{\mathcal{L}}(s,y;t,x)[\widehat\xi] = \\
   \ds \qquad = -\int_s^t U(\tau,\widehat\xi(\tau))\,d\tau
   \le (t-s)\left(-\min_{(t,x) \in \rset^2} U(t,x)\right).
\end{array}
$$
Multiplying these inequalities by $(t-s)^{-1}$, we obtain~(\ref{updown}).
\hfill$\square$

\begin{lm}
Let $R$, $x_1$, $x_2$, $y_1$, $y_2$, $s$, $t \in \rset$ be such that $s <
t$, $R > 0$, $|x_i - y_j| \le R(t-s)$, $i$,~$j = 1,2$; then there exists
$C = C(s,t,R) \in \rset$, $C > 0$, such that
\begin{equation}
   |L(s,y_1;t,x_1) - L(s,y_2;t,x_2)| \le C(|y_1 - y_2| + |x_1 - x_2|).
\label{Lip}
\end{equation}
\label{l:Lip}
\end{lm}
In other words, $L(s,y;t,x)$ is locally Lipschitzian.
\smallskip
\par\noindent\textsc{Proof.} It is sufficient to prove the inequality
\begin{equation}
   |L(s,y;t,x_1) - L(s,y;t,x_2)| \le C|x_1 - x_2|,
\label{temp1}
\end{equation}
where $|x_i - y| \le R(t-s)$, $i = 1,2$. The inequality for the first pair
of arguments of the function $L$ is proved similarly; combining these two
inequalities, we obtain~(\ref{Lip}).

Let $\xi_0 \in \Omega(s,y;t,x_1)$ be a minimizer of the functional
$\mathcal{L}(s,y;t,x_1)$. By lemma~\ref{l:Cesari}, it follows that
$|\xi_0'(\tau)| \le C_1 = C_1(R,s,t)$, $s \le \tau \le t$.
Consider the trajectory $\xi(\tau) = \xi_0(\tau) + (x_2 - x_1)(\tau - s) /
(t-s)$, $s \le \tau \le t$. It is clear that $\xi \in \Omega(s,y;t,x_2)$.
We have
$$
\begin{array}{l}
   \ds L(s,y;t,x_2) - L(s,y;t,x_1)
   \le \mathcal{L}(s,y;t,x_2)[\xi] - \mathcal{L}(s,y;t,x_1)[\xi_0] = \\
   \ds\qquad = \int_s^t \left[L_0\left(\xi'_0(\tau)
   + {x_2 - x_1\over t-s}\right) - L_0(\xi'_0(\tau))\right]\,d\tau - \\
   \ds\qquad\quad- \int_s^t \left[U\left(\tau, \xi_0(\tau)
   + {x_2 - x_1\over t-s}(\tau - s)\right)
   - U(\tau,\xi_0(\tau))\right]\,d\tau.
\end{array}
$$
The functions $U(t,x)$ and $L_0(v)$ are of class $C^1$; in addition, it
follows from~$(H_3)$ and $(H_4)$ that $|\partial U(t,x)/\partial x| \le
M_1$ for some positive $M_1 \in \rset$. Hence,
$$
\begin{array}{l}
   \ds L(s,y;t,x_2) - L(s,y;t,x_1) \le \\
   \ds\qquad\le {\ds|x_2 - x_1|\over \ds t - s}(t - s)
   \max_{\tau \in [s,t], |\theta| \le 1}
   \left|L'_0\left(\xi'_0(\tau) + \theta{|x_2 - x_1|\over t-s}\right)\right|
   + \\
   \ds\qquad\quad + {\ds|x_2 - x_1|\over \ds t-s}{(t-s)^2\over 2}M_1.
\end{array}
$$
We recall that $L'_0(v)$ is finite for all $v$; thus for some $C =
C(s,t,R,M_1) \in \rset$, $C > 0$,
$$
   L(s,y;t,x_2) - L(s,y;t,x_1) \le C|x_2 - x_1|.
$$
Similarly, $L(s,y;t,x_2) - L(s,y;t,x_1) \ge C|x_2 - x_1|$. This completes
the proof of inequality~(\ref{temp1}). \hfill$\square$

We see that $L(s,y;t,x)$ is continuous in $x$ and~$y$; therefore it follows
from lemma~\ref{l:updown} that infimum is attained in the Lax--Ole\u{\i}nik
formula~(\ref{FLO}). In the sequel, we shall use $\max$ instead of $\inf$
in all formulas derived from~(\ref{FLO}).

\begin{prop}
Let $x_1$, $x_2$, $y_1$, $y_2$, $s$, $r$, $t \in \rset$, $s < r < t$;
suppose there exists $z_0 \in \rset$ such that for all $z \in \rset$
\begin{equation}
   L(s,y_i;r,z) + L(r,z;t,x_i) \ge L(s,y_i;r,z_0) + L(r,z_0;t,x_i),\quad
   i = 1,2.
\label{temp2}
\end{equation}
Then either $x_1 = x_2$ and $y_1 = y_2$ or $(x_1 - x_2)(y_1 - y_2) < 0$.
\label{p:Bangert}
\end{prop}
\textsc{Proof.} For $x_1 = x_2$ and $y_1 = y_2$ there is nothing to prove.
Assume $x_1 \neq x_2$ or $y_1 \neq y_2$. Let $\xi_i^< \in
\Omega(s,y_i;r,z_0)$ and $\xi_i^> \in \Omega(r,z_0;t,x_i)$ be minimizers of
functionals $\mathcal{L}(s,y_i;r,z_0)$ and $\mathcal{L}(r,z_0;t,x_i)$, $i =
1,2$. By~(\ref{temp2}) and lemma~\ref{l:Bellman}, it follows that for $i =
1,2$
$$
   L(s,y_i;r,z_0) + L(r,z_0;t,x_i) = L(s,y_i;t,x_i).
$$
Hence for $i = 1,2$ the trajectories
$$
   \xi_i(\tau) = \left\{
   \begin{array}{ll}
      \xi_i^<(\tau), &s \le \tau \le r \\
      \xi_i^>(\tau), &r \le \tau \le s
   \end{array}\right.
   \quad(\xi_i \in \Omega(s,y_i;t,x_i))
$$
are minimizers of $\mathcal{L}(s,y_i;t,x_i)$. Using
lemma~\ref{l:Cesari}, we see that the trajectories $\xi_i$, $i =
1,2$, are of class $C^1$. This means that $(\xi^<_i)'(r-0) =
(\xi^>_i)'(r+0) = \xi'_i(r)$, $i = 1,2$.

We claim that $\xi'_1(r) \neq \xi'_2(r)$. Indeed, for $i = 1,2$ let
$\lambda_i$ be the $C^1$ functions corresponding to trajectories $\xi_i$ by
lemma~\ref{l:Cesari}. Assume that $\xi'_1(r) = \xi'_2(r)$; then, by
continuity of $L'_0(\cdot)$, it follows from~(\ref{Euler}) that
$\lambda_1(r) = \lambda_2(r)$. Consider the Cauchy problem for the system
of ordinary differential equations~(\ref{Euler}) on the interval $[s,t]$
with initial data $\xi_0(r) = \xi_1(r) = \xi_2(r)$ and $\lambda(r) =
\lambda_1(r) = \lambda_2(r)$. It follows from assumption $(H_3)$ that the
function $\partial U(t,x)/\partial x$ is Lipschitzian and its absolute
value is bounded by the constant $M_1 > 0$; thus it follows from the second
equation~(\ref{Euler}) that $|\lambda(\tau) - \lambda(r)| \le M_1(t - s)$
for all $\tau \in [s,t]$. Hence it follows from assumption $(H_1)$ that the
system
$$
   \xi'_0(\tau) = H'_0(\lambda(\tau)), \qquad
   \lambda'(\tau) = - \pd{U(\tau, \xi_0(\tau)}{\xi_0}, \qquad
   s < \tau < t,
$$
which is equivalent to~(\ref{Euler}), satisfies the conditions of the
uniqueness theorem. Thus $\xi_1(\tau) = \xi_2(\tau)$, $s \le \tau \le t$,
and in particular $x_1 = x_2$ and $y_1 = y_2$. This contradiction proves
that $\xi'_1(r) \neq \xi'_2(r)$.

Now we prove that there exists $\delta > 0$ such that $s < r - \delta$, $r
+ \delta < t$, and if $\tau \in (r-\delta, r+\delta)$, then $\xi_1(\tau)$
and $\xi_2(\tau)$ coincide only at $\tau = r$. Indeed, let a sequence
$\{t_n\}$ be such that for all $n$, $s < t_n < t$, $t_n \neq r$,
$\xi_1(t_n) = \xi_2(t_n)$, and $\lim_{n \to \infty} t_n = r$. This implies
that
$$
   \xi'_1(r)
   = \lim_{n \to \infty}{\xi_1(t_n) - \xi_1(r)\over t_n - r}
   = \lim_{n \to \infty}{\xi_2(t_n) - \xi_2(r)\over t_n - r}
   = \xi'_2(r),
$$
a contradiction.

Now assume that $(x_1 - x_2)(y_1 - y_2) \ge 0$; to be precise, let $x_1 \le
x_2$ and $y_1 \le y_2$. For $i = 1,2$ define trajectories $\bar\xi_i \in
\Omega(s,y_i;t,x_i)$ by $\bar\xi_1(\tau) = \min\{\xi_1(\tau),
\xi_2(\tau)\}$ and $\bar\xi_2(\tau) = \max\{\xi_1(\tau), \xi_2(\tau)\}$. By
the above each of $\bar\xi_1$ and $\bar\xi_2$ coincides with $\xi_1$ and
$\xi_2$ on a finite number of intervals of finite length. Let us check
that the trajectories $\bar\xi_i$, $i = 1,2$, are minimizers. Indeed, we
have
$$
   \mathcal{L}(s,y_i;t,x_i)[\bar\xi_i]
   \ge \mathcal{L}(s,y_i;t,x_i)[\xi_i]
   = L(s,y_i;t,x_i),
   \qquad i = 1,2,
$$
and
$$
\begin{array}{l}
   \mathcal{L}(s,y_1;t,x_1)[\bar\xi_1]
   + \mathcal{L}(s,y_2;t,x_2)[\bar\xi_2] = \\
   \quad = \mathcal{L}(s,y_1;t,x_1)[\xi_1]
   + \mathcal{L}(s,y_2;t,x_2)[\xi_2] = \\
   \quad = L(s,y_1;t,x_1) + L(s,y_2;t,x_2).
\end{array}
$$
Therefore $\mathcal{L}(s,y_i;t,x_i)[\bar\xi_i] = L(s,y_i;t,x_i)$, $i =
1,2$.

By lemma~\ref{l:Cesari}, the trajectories $\bar\xi_i$, $i = 1,2$, are
of class $C^1$. This implies that $\bar\xi'_i(r-0) = \bar\xi'_i(r+0)$, $i =
1,2$, where all one-sided derivatives exist. But this means that $\xi'_1(r
- 0) = \xi'_2(r + 0)$ and $\xi'_2(r - 0) = \xi'_1(r + 0)$, since $r$ is the
only intersection point of $\xi_1$ and $\xi_2$ in the interval $(r -
\delta, r + \delta)$. Thus $\xi'_1(r) = \xi'_2(r)$. This contradiction
concludes the proof. \hfill$\square$
\medskip
\par\noindent\textbf{Corollary} \emph{Let $x_1$, $x_2$, $y_1$, $y_2$, $s$,
$t \in \rset$ be such that $x_1 < x_2$, $y_1 < y_2$, and $s < t$. Then}
\begin{equation}
   L(s,y_1;t,x_1) + L(s,y_2;t,x_2)
   < L(s,y_2;t,x_1) + L(s,y_1;t,x_2).
\label{twist}
\end{equation}
\medskip
\par\noindent\textsc{Proof.} Let $\xi_1 \in \Omega(s,y_1;t,x_2)$ and $\xi_2
\in \Omega(s,y_2;t,x_1)$ be minimizers of functionals
$\mathcal{L}(s,y_1;t,x_2)$ and $\mathcal{L}(s,y_2;t,x_1)$, respectively.
Note that $\xi_1(s) < \xi_2(s)$ and $\xi_1(t) > \xi_2(t)$; hence there
exist $r$, $z_0$, $s < r < t$, such that $\xi_1(r) = \xi_2(r) = z_0$.
Therefore $L(s,y_1;r,z_0) + L(r,z_0;t,x_2) = L(s,y_1;t,x_2)$,
$L(s,y_2;r,z_0) + L(r,z_0;t,x_1) = L(s,y_2;t,x_1)$ and by
lemma~\ref{l:Bellman} $ L(s,y_1;r,z_0) + L(r,z_0;t,x_1) \ge
L(s,y_1;t,x_1)$, $L(s,y_2;r,z_0) + L(r,z_0;t,x_2) \ge L(s,y_2;t,x_2)$. This
implies that $L(s,y_1;t,x_1) + L(s,y_2;t,x_2) \le L(s,y_1;t,x_2) +
L(s,y_2;t,x_1)$.  Assume that this weak inequality is actually an equality.
This assumption means that for all $z \in \rset$
$$
\begin{array}{c}
   \ds L(s,y_1;t,x_1)
   = L(s,y_1;r,z_0) + L(r,z_0;t,x_1)
   \le L(s,y_1;r,z) + L(r,z;t,x_1), \\
   \ds L(s,y_2;t,x_2)
   = L(s,y_2;r,z_0) + L(r,z_0;t,x_2)
   \le L(s,y_2;r,z) + L(r,z;t,x_2).
\end{array}
$$
It follows from the previous lemma that $(x_1 - x_2)(y_1 - y_2) < 0$.
This contradiction concludes the proof. \hfill$\square$

\section{Reduction to a functional equation}

Let $a \in \rset$, $s_0(x)$ be a continuous periodic function. Suppose
$S(t,x)$ is the solution~(\ref{FLO}) of the Cauchy problem for
equation~(\ref{HJE}) with the initial function $S_0(x) = ax + s_0(x)$.
Since the Lagrangian $L(t,x,v) = L_0(v) - U(t,x)$ is periodic in $t$, it is
natural to consider $S(t,x)$ at integer values of $t$: $t = n = 1,2,\ldots$
Using~(\ref{FLO}), we get
$$
   S(n,x) - ax = \min_{y \in \rset}\,(s_0(y) + L(0,y;n,x) + a(y - x))
   = s_n(x;a,s_0).
$$
As a function of $x$, $s_n(x;a,s_0)$ is a pointwise minimum of a family of
continuous functions, so it is continuous. Further, it follows from the
corollary to lemma~\ref{l:diagper} that it is periodic; thus $s_n(x;a,s_0)
= \min_{k \in \zset}\,s_n(x - k; a,s_0)$. Using the above formula,
lemma~\ref{l:diagper}, and periodicity of $s_0(y)$, we get
$$
   s_n(x;a,s_0) = \min_{k \in \zset} \min_{y \in \rset}\,
   (s_0(y) + L(0,y + k;n,x) + a(y + k - x)).
$$
Thus for any $n > 0$ and $y$, $x \in \rset$ the quantity $L(0,y+k;n,x) +
a(y + k - x)$ is bounded from below as a function of $k \in \zset$. We
denote
\begin{equation}
   L^a_n(y,x) = \min_{k \in \zset}\,(L(0,y + k;n,x) + a(y + k - x)),
\label{defLan}
\end{equation}
where minimum is attained since $L(s,y;t,x)$ is continuous and grows
superlinearly as $|x - y| \to \infty$; hence,
\begin{equation}
   s_n(x;a,s_0) = \min_{y \in \rset}\,(s_0(y) + L^a_n(y,x)).
\label{defsn}
\end{equation}
\begin{prop}
For any $n = 1,2,\ldots$, the function $L^a_n(y,x)$ is periodic and
Lipschitzian in $x$; the Lipschitz constant $C^*(a)$ does not depend on
$n$ and $y \in \rset$. For any $n_1 > 0$, $n_2 > 0$,
\begin{equation}
   L^a_{n_1 + n_2}(y,x)
   = \min_{z \in \rset}\,(L^a_{n_1}(y,z) + L^a_{n_2}(z,x)).
\label{Lan1n2}
\end{equation}
\label{p:LanLip}
\end{prop}
\textsc{Proof.} Using the definition~(\ref{defLan}) and
lemma~\ref{l:diagper}, we obtain $L^a_n(y + l,x + m) = L^a_n(y,x)$ for any
integer $l$, $m$; thus $L^a_n(y,x)$ is periodic. Combining~(\ref{defLan})
for $n = n_1 + n_2$ with (\ref{Bellman}) for $s = 0$, $r = n_1$, and $t
= n_1 + n_2$ and using lemma~\ref{l:diagper}, we get
$$
   L^a_{n_1+n_2}(y,x)
   = \min_{z \in \rset}\,(L^a_{n_1}(y,z) + L(0,z;n_2,x) + a(z-x)).
$$
Using periodicity of $L^a_{n_1}(y,z)$ in $z$, we obtain~(\ref{Lan1n2}).

Let us prove that $L^a_1(y,x)$ is Lipschitzian. Since $L^a_1(y,x)$ is a
pointwise minimum of a family of continuous functions, it is
continuous. Denote by $N$ the maximum value of $L^a_1(y,x)$ on the set $Q =
\{\, (y,x) \mid 0 \le y \le 1, 0 \le x \le 1 \,\}$. Clearly, there exists
$R > 0$ such that $L_0(x - y - k) + a(y + k - x) > N + \max_{(t,x) \in
\rset^2}\,U(t,x)$ for all $(y,x) \in Q$ whenever $|k| > R$. Thus it follows
from lemma~\ref{l:updown} that for any $(y,x) \in Q$
$$
   L^a_1(y,x) = \min_{k \in \zset, |k| < R}\,
   (L(0,y + k;1,x) + a(y + k - x)).
$$
On the other hand, using lemma~\ref{l:Lip}, we see that all functions
$L(0,y + k;1,x) + a(y + k - x)$, $|k| < R$, are Lipschitzian on $Q$
with the constant $C^*(a) = C(0,1,R+1) + |a|$. Thus $L^a_1(y,x)$ is
Lipschitzian on $Q$ (and, by periodicity, on the whole $\rset^2$) with the
same constant.

Now it follows from~(\ref{Lan1n2}) that for any $n \ge 2$
$$
   L^a_n(y,x) = \min_{z \in \rset}\,(L^a_{n-1}(y,z) + L^a_1(z,x)),
$$
that is for any $y \in \rset$ $L^a_n(y,x)$ as a function of $x$ is a
pointwise minimum of a family of functions sharing the same Lipschitz
constant $C^*(a)$. Thus $L^a_n(y,x)$ is Lipschitz continuous in $x$ with
the constant $C^*(a)$ for all $n = 1,2,\ldots$
\hfill$\square$
\medskip
\par\noindent\textbf{Corollary} \emph{For any $n = 1,2,\ldots$ the function
$s_n(x;a,s_0)$ defined in~(\ref{defsn}) is Lipschitzian with
the constant $C^*(a)$; it satisfies}
\begin{equation}
   \max_{x \in \rset}\,s_n(x;a,s_0)
   - \min_{x \in \rset}\,s_n(x;a,s_0)
   \le C^*(a)
\label{oscsn}
\end{equation}
\emph{and for all $m \in \zset$, $m \ge n$,}
\begin{equation}
   s_m(x;a,s_0)
   = \min_{y \in \rset}\,(s_{m-n}(y;a,s_0) + L^a_n(y,x)).
\label{smsn}
\end{equation}
\medskip
\par\noindent\textsc{Proof.} From~(\ref{defsn}) it follows that
$s_n(x;a,s_0)$ is Lipschitzian with the constant $C^*(a)$ as a pointwise
minimum of a family of $C^*(a)$-Lipschitzian functions. Taking into account
periodicity of $s_n(x;a,s_0)$, we obtain~(\ref{oscsn}).
Equation~(\ref{smsn}) follows from~(\ref{Lan1n2}) for $n_1 = m - n$, $n_2 =
n$. \hfill$\square$

Suppose $S^a(t,x)$ is a viscosity solution to equation~(\ref{HJE}) of the
form~(\ref{defSa}); then for $t > 0$ it is a solution of Cauchy problem for
equation~(\ref{HJE}) with the initial data $S_0(x) = S^a(0,x)$. Hence,
$$
   S^a(t,x) = \min_{y \in \rset}\,(S^a(0,y) + L(0,y;t,x)),\quad t>0.
$$
Denote $s^a(x) = S^a(0,x) - ax = s^a(0,x)$. It follows from~(\ref{defSa})
that $S^a(1,x) - ax + H(a) = s^a(x)$. Thus,
$$
   \min_{y \in \rset}\,(s^a(y) + a(y - x) + L(0,y;1,x)) + H(a) = s^a(x).
$$
Using the above notation, we rewrite this as
\begin{equation}
   s^a(x) = \min_{y \in \rset}\,(s^a(y) + L^a_1(y,x)) + H(a).
\label{functeq}
\end{equation}

On the other hand, let $s^a(x)$ satisfy the functional
equation~(\ref{functeq}); then the solution~(\ref{FLO}) of the Cauchy
problem for equation~(\ref{HJE}) with the initial data $S_0(x) = ax +
s^a(x)$ has the form~(\ref{defSa}). Thus to prove theorem~I it is
sufficient to show that for any $a \in \rset$ there exist a number $H(a)$
and a continuous periodic function $s^a(x)$ such that~(\ref{functeq}) is
satisfied.

\section{Proof of theorem~I}

Let $a \in \rset$. We denote
\begin{equation}
   H_n(a)
   = \max_{(y,x) \in \rset^2}\,\left(-{1\over n}L^a_n(y,x)\right)
   = \max_{(y,x) \in \rset^2}\,
   \left(-{1\over n}L(0,y;n,x) + a{x-y\over n}\right).
\label{defHn}
\end{equation}
\begin{prop}
For any $a \in \rset$ there exists $H(a) \in \rset$ such that
\begin{equation}
   |H_n(a) - H(a)| \le {C^*(a)\over n},
\label{HnHa}
\end{equation}
where $C^*(a)$ is the Lipschitz constant of the function $L^a_n(y,x)$. The
function $a \mapsto H(a)$ is convex and satisfies inequalities
\begin{equation}
   m \le H(a) - H_0(a) \le M,
\label{estHH0rep}
\end{equation}
where $m$ and $M$ were defined in lemma~\ref{l:updown}.
\label{p:Ha}
\end{prop}
\textsc{Proof.} Let $s_0(x) = 0$, $s_n(x) = s_n(x;a,s_0)$
(see~(\ref{defsn})). It follows from~(\ref{smsn}) that for any integer
$n$, $n_0$, $0 < n_0 < n$,
\begin{equation}
   s_{n}(x) = \min_{y \in \rset}\,(s_{n - n_0}(y) + L^a_{n_0}(y,x)).
\label{temp3}
\end{equation}
We see that $\min_{x \in \rset}\,s_n(x) = \min_{(y,x) \in \rset^2}\,
L^a_n(y,x) = -n H_n(a)$ for any $n = 1,2,\ldots$ It follows from the
corollary to proposition~\ref{p:LanLip} that $-nH_n(a) \le s_n(x) \le
-nH_n(a) + C^*(a)$. Combining these inequalities with~(\ref{temp3})
and~(\ref{defHn}), we obtain
$$
\begin{array}{l}
   -nH_n(a)
   \le s_n(x)
   \le -(n - n_0)H_{n - n_0}(a) + C^*(a) - n_0 H_{n_0}(a),\\
   -(n - n_0)H_{n - n_0}(a) - n_0 H_{n_0}(a)
   \le s_n(x)
   \le -nH_n(a) + C^*(a).
\end{array}
$$
Hence,
$$
   |nH_n(a) - n_0H_{n_0}(a) - (n - n_0)H_{n - n_0}(a)| \le C^*(a).
$$
Since $n > n_0$, there exist integer $p \ge 1$ and $0 \le q < n$ such that
$n = pn_0 + q$. Then by induction over $p$ it is easily checked that $|n
H_n(a) - pn_0 H_{n_0}(a) - qH_q(a)| \le pC^*(a)$, that is
$$
   \left| H_n(a) - H_{n_0}(a) + {q\over n}(H_{n_0}(a) - H_q(a))\right|
   \le {p\over n}C^*(a) \le {1\over n_0}C^*(a).
$$
Let $\varepsilon > 0$ be arbitrarily small, $n_0$ be so large that the
right-hand side of this inequality is less than $\varepsilon$, and $N$ be
such that $(1/N)\max_{1 \le q \le n_0}\, |q(H_{n_0}(a) - H_q(a)| <
\varepsilon$; then $|H_{n_1}(a) - H_{n_2}(a)| < 4\varepsilon$ for all
$n_1$, $n_2 \ge N$. Thus $\{H_n(a)\}$, $n = 1,2,\ldots$, is a Cauchy
sequence. Denote its limit by $H(a)$; then we get~(\ref{HnHa}) from the
above inequality in the limit $n \to \infty$.

Suppose $a_1$,~$a_2 \in \rset$, $0 < \alpha, \beta < 1$, $\alpha + \beta =
1$. Using definition of $H_n(a)$~(\ref{defHn}), we obtain
$$
\begin{array}{l}
   H_n(\alpha a_1 + \beta a_2)
   = -{1\over n} \min_{(y,x) \in \rset^2}\,
   (L(0,y;n,x) + (\alpha a_1 + \beta a_2)(y - x)) = \\
   \quad = -{1\over n} \min_{(y,x) \in \rset^2}\,
   [\alpha(L(0,y;n,x) + a_1(y - x)) + \beta(L(0,y;n,x) + a(y - x))) \le \\
   \quad \le \alpha H_n(a_1) + \beta H_n(a_2).
\end{array}
$$
In the limit $n \to \infty$ this implies that the function $H(a)$ is
convex.

Finally, lemma~\ref{l:updown} implies that $-M \le (1/n)L(0,y;n,x) -
L_0((x-y)/n) \le -m$. Hence,
$$
   m + a{x-y\over n} - L_0\left({x-y\over n}\right)
   \le a{x-y\over n} - {1\over n}L(0,y;n,x)
   \le M + a{x-y\over n} - L_0\left({x-y\over n}\right).
$$
Taking $\max$ over $(y,x) \in \rset^2$, using~(\ref{defHn}), and denoting
$(x-y)/n$ by $v$, we obtain
$$
   m + \max_{v \in \rset}\,(av - L_0(v))
   \le H_n(a)
   \le M + \max_{v \in \rset}\,(av - L_0(v)).
$$
But by a well-known formula for the Legendre transform $\max_{v \in
\rset}\,(av - L_0(v)) = H_0(a)$. Thus we obtain~(\ref{estHH0rep}) in the
limit $n \to \infty$. \hfill$\square$

Suppose $a \in \rset$ and $s_0(x)$ is a continuous and periodic function.
We claim that there exists
\begin{equation}
   \liminf_{n \to \infty}\,(s_n(x;a,s_0) + nH(a)) = s^a(x)
\label{saliminf}
\end{equation}
and $s^a(x)$ satisfies~(\ref{functeq}), i.e., the solution of the Cauchy
problem for equation~(\ref{HJE}) with the initial data $S_0(x) = ax +
s^a(x)$ has the form~(\ref{defSa}). Combining this with
proposition~\ref{p:Ha}, we get the statement of theorem~I.

Using definitions of $s_n(x;a,s_0)$ and $H_n(a)$, we get
$$
   \min_{y \in \rset}\,s_0(y) - nH_n(a)
   \le s_n(x;a,s_0)
   \le \max_{y \in \rset}\,s_0(y) - nH_n(a).
$$
Adding $nH(a)$ and using~(\ref{HnHa}), we obtain
\begin{equation}
   \min_{y \in \rset}\,s_0(y) - C^*(a)
   \le s_n(x;a,s_0) + nH(a)
   \le \max_{y \in \rset}\,s_0(y) + C^*(a).
\label{temp4}
\end{equation}
Let
$$
   \bar s_n(x) = \inf_{m \ge n}\,(s_m(x;a,s_0) + mH(a)),\quad
   n = 1,2,\ldots.
$$
For any $x \in \rset$ the sequence $\{\bar s_n(x)\}$ is nondecreasing; it
follows from~(\ref{temp4}) that it is bounded. Further, the corollary to
proposition~\ref{p:LanLip} implies that all functions $\bar s_n(x)$ are
Lipschitzian with the constant $C^*(a)$; hence, this sequence is
equicontinuous. It follows that there exists
$$
   \lim_{n \to \infty} \bar s_n(x)
   = \liminf_{n \to \infty}\,(s_n(x;a,s_0) + nH(a))
   = s^a(x)
$$
and $s^a(x)$ is periodic and continuous.

Let us check that $s^a(x)$ satisfies~(\ref{functeq}). We have
$$
\begin{array}{l}
   \ds\min_{y \in \rset}\,(\bar s_n(y) + L^a_1(y,x)) + H(a) = \\
   \ds\quad = \min_{y \in \rset}\,(\inf_{m \ge n}\,(\min_{z \in \rset}\,
   (s_0(z) + L^a_m(z,y) + L^a_1(y,x) + (m+1)H(a)))).
\end{array}
$$
But it follows from the definition of $H_n(a)$ and proposition~\ref{p:Ha}
that $L^a_m(z,y) + mH(a) \ge -C^*(a)$ for all $m = 1,2,\ldots$ Therefore we
can take minimum over $y \in \rset$ before infimum over $m \ge n$ and
obtain
$$
   \min_{y \in \rset}\,(\bar s_n(y) + L^a_1(y,x)) + H(a) =
   \bar s_{n+1}(x).
$$
After passage to the limit $n \to \infty$ this yields that $s^a(x)$
satisfies~(\ref{functeq}).

\section{A many-valued map associated with $s^a(x)$}

Suppose $a \in \rset$, $s^a(x)$ is a continuous periodic function
satisfying~(\ref{functeq}). Denote $L_a(y,x) = L(0,y;1,x) + a(y-x) + H(a)$.
Let $Y^a(x)$ be the many-valued map of $\rset$ to $\rset$ defined by
\begin{equation}
   Y^a(x) = \arg \min_{y \in \rset}\,(s^a(y) + L_a(y,x)),
\label{defYa}
\end{equation}
where
$$
   \arg \min_{z \in Z}f(z)
   = \{\, z \in Z \mid f(t) \ge f(z) \mbox{ for all $t \in Z$} \,\}.
$$
It is readily seen that if $x$, $y \in \rset$, then $y \in Y^a(x)$ iff
\begin{equation}
   s^a(y) + L_a(y,x) = s^a(x)
   = \min_{z \in \rset}\,(s^a(z) + L_a(z,x)).
\label{ncc}
\end{equation}

For any $X \subset \rset$, $Y \subset \rset$, by $X + Y$ we denote the set
$\{z \in \rset|\, z = x + y, x \in X, y \in Y\}$. If $Y = \{y\}$, by $X + y$
we mean $X + \{y\}$.
\begin{lm}
For all $x \in \rset$
\begin{equation}
   Y^a(x+1) = Y^a(x) + 1.
\label{Yper}
\end{equation}
If there exists $x_0 \in \rset$ such that $x \in Y^a(x_0)$, then the set
$Y^a(x)$ contains only one point. If $x_1$, $x_2$, $y_1$, $y_2 \in \rset$,
$x_1 < x_2$, $y_1 \in Y^a(x_1)$, and $y_2 \in Y^a(x_2)$, then $y_1 < y_2$.
\label{l:Yper}
\end{lm}
\textsc{Proof.} Equation~(\ref{Yper}) follows from periodicity of $s^a(x)$
and lemma~\ref{l:diagper}.

Suppose $x_1 < x_2$ and $y_i \in Y^a(x_i)$, $i = 1,2$. By~(\ref{ncc}), it
follows that
$$
\begin{array}{l}
   s^a(y_1) + L(0,y_1;1,x_1) + ay_1 \le s^a(y_2) + L(0,y_2;1,x_1) + ay_2, \\
   s^a(y_2) + L(0,y_2;1,x_2) + ay_2 \le s^a(y_1) + L(0,y_1;1,x_2) + ay_1,
\end{array}
$$
that is
$$
   L(0,y_1;1,x_1) + L(0,y_2;1,x_2) \le L(0,y_2;1,x_1) + L(0,y_1;1,x_2).
$$
Thus it follows from the corollary to proposition~\ref{p:Bangert} that $y_1
\le y_2$.

Let us check that $y_1 < y_2$. Assume the converse; let $y = y_1 = y_2$.
Take any $y_0 \in Y^a(y)$. By~(\ref{ncc}), it follows that
$$
   s^a(x_i) = s^a(y_0) + L(0,y_0;1,y) + L(1,y;2,x_i) + a(y_0 - x_i) + 2H(a),
   \quad i = 1,2.
$$
On the other hand, using~(\ref{functeq}), we get
$$
\begin{array}{l}
   \ds s^a(x_i) = \min_{z \in \rset} \min_{t \in \rset}\,
   (s^a(t) + L(0,t;1,z) + L(1,z;2,x_i) + a(t - x_i) + 2H(a)) \le \\
   \ds\qquad \le \min_{z \in \rset}\,
   (s^a(y_0) + L(0,y_0;1,z) + L(1,z;2,x_i) + a(y_0 - x_i) + 2H(a)), \\
   \qquad i = 1,2.
\end{array}
$$
Hence it follows from proposition~\ref{p:Bangert} that $x_1 = x_2$; this
contradiction proves that $y_1 < y_2$.

Finally, suppose $x \in Y^a(x_0)$ for some $x_0 \in \rset$ and $y_1$, $y_2
\in Y^a(x)$. By a similar argument, we see that $y_1 = y_2$. Thus
$Y^a(x)$ consists of a single point. \hfill$\square$

\begin{prop}
Suppose $M_0 = \rset$, $M_n = Y^a(M_{n-1})$, $n = 1,2,\ldots$; then $M_0
\supset M_1 \supset M_2 \supset \ldots$ and there exists a nonempty closed
set $M^a$ such that (i) $M^a = \bigcap_{n = 0}^\infty M_n$; (ii) for any
open $U \subset \rset$ such that $M^a \subset U + \zset$ there exists $N >
0$ with the following property: if the sequence $\{y_n\}$ is such that
$y_{n+1} \in Y^a(y_n)$, $n = 0,1,\ldots$, then $y_n \in U + \zset$ for all
$n \ge N$; (iii) the restriction $Y^a|_{M^a}$ of $Y^a$ to $M^a$ is a
single-valued bijective continuous map of the set $M^a$ into itself with
continuous inverse.
\label{p:Ma}
\end{prop}
\textsc{Proof.} Evidently, all $M_n \neq \varnothing$ and $M_0 = \rset$
contains $M_1$. Suppose $M_n \subset M_{n-1}$ for some $n \ge 1$, $x \in
M_n$; then $Y^a(x) \subset Y^a(M_n) \subset Y^a(M_{n-1}) = M_n$.  Hence,
$M_{n+1} = Y^a(M_n) \subset M_n$.

Let us show that all $M_n$, $n = 1,2,\ldots$, are closed. Note that $M_0 =
\rset$ is closed. Suppose $M_n$ is closed and the sequence $y_k \in
M_{n+1}$, $k = 1,2,\ldots$, has a limit $\bar y \in \rset$. It follows from
lemma~\ref{l:Yper} that for any $k$ there exists a unique $x_k =
(Y^a)^{-1}(y_k) \in M_n$ and all $x_k$ are contained in a bounded interval.
Thus there exists a subsequence $\{x_{k_l}\}$ that converges to a limit
$\bar x \in M_n$. Using~(\ref{ncc}), we get
$$
   s^a(y_{k_l}) + L_a(y_{k_l},x_{k_l}) = s^a(x_{k_l}).
$$
Since the finctions $s^a(x)$ and $L_a(y,x)$ are continuous, we obtain
$$
   s^a(\bar y) + L_a(\bar y,\bar x) = s^a(\bar x).
$$
Now it follows from~(\ref{ncc}) that $\bar y \in Y^a(\bar x)$. Hence $\bar
y \in Y^a(M_n) = M_{n+1}$, that is $M_{n+1}$ is closed.

Consider the topology $\mathcal{T}$ on $\rset$ such that $V \in \mathcal{T}$
iff $V = U + \zset$, where $U$ is open in the usual sense. Note that
$\rset$ is compact in this topology. It follows from~(\ref{Yper}) that
complements $U_n$ of sets $M_n$ are open in the topology $\mathcal{T}$.
We see that $U_n \subset U_{n+1}$.

Clearly, the set $M^a = \bigcap_{n=0}^\infty M_n$ is closed. Let us check
that it is nonempty. Assume the converse; hence sets $U_n$, $n =
1,2,\ldots$ cover all $\rset$. Since $\rset$ is compact in the topology
$\mathcal{T}$, we see that there exists $N > 0$ such that $\rset =
\bigcup_{n = 0}^N U_n = U_N$. Thus $M_N = \varnothing$; this contradiction
proves that $M^a$ is nonempty.

Now let $V \in \mathcal{T}$ be such that $M^a \subset V$. Arguing as above,
we obtain that there exists $N > 0$ such that $\rset = V \cup U_N$. On the
other hand, for any sequence $\{y_n\}$ such that $y_{n+1} \in Y^a(y_n)$, $n
= 0,1,\ldots$, it follows that $y_n \in M_n$. Thus for all $n \ge N$ we
obtain $y_n \in V$.

Denote by $Y^a|_{M_n}$:~$M_n \to M_{n+1}$ the restriction of $Y^a$ to
$M_n$, $n = 1,2,\ldots$ It follows from lemma~\ref{l:Yper} that
$Y^a|_{M_n}$ is single-valued, strictly increasing as a function $\rset
\to \rset$, and bijective. We claim that it is a homeomorphism. Indeed,
let sequences $\{x_k\} \subset M_n$ and $\{y_k\} \subset M_{n+1}$ be such
that $Y^a(x_k) = y_k$ for all $k$. If $y_k$ converge to $\bar y$, then it
follows from strict monotonicity of $Y^a|_{M_n}$ that $x_k$ converge to a
unique $\bar x$; hence the map $Y^a|_{M_n}$ has a continuous inverse.
Further, if $x_k$ converge to $\bar x$, then the set $\{y_k\}$ is bounded.
Let $\bar y_1$, $\bar y_2$, $y_1 \neq y_2$, be two limit points of
$\{y_k\}$; then we see from~(\ref{ncc}) that $\{\bar y_1,\bar y_2\} \subset
Y^a(\bar x)$, since $s^a(x)$ and $L_a(y,x)$ are continuous. But the map
$Y^a|_{M_n}$ is single-valued; this contradiction proves that it is
continuous.

By the above $M^a$ is a closed subset of $M_n$ for all $n = 1,2\ldots$;
in addition, $Y^a(M^a) = M^a$. It follows that the restriction of
$Y^a|_{M_n}$ to $M^a$ is an (auto)homeomorphism of the set $M^a$.
\hfill$\square$

We say that $M^a$ is \emph{the invariant set} of the many-valued map
$Y^a(x)$.

\section{The connection with Aubry--Mather theory}

The following remarks concern the results of Aubry--Mather theory that are
needed for the proof of theorem~II.

Let a function $L(y,x)$ be everywhere finite and continuous on $\rset^2$,
diagonally periodic in the sense of lemma~\ref{l:diagper}, tend to plus
infinity as $|x-y| \to \infty$, and for all $x_1$, $x_2$, $y_1$, $y_2 \in
\rset$ satisfy the following conditions: (i) $L(y_1,x_1) + L(y_2,x_2) \le
L(y_2,x_1) + L(y_1,x_2)$ whenever $x_1 < x_2$, $y_1 < y_2$; (ii) if $z_0
\in \rset$ is such that for all $z \in \rset$ $L(y_i,z) + L(z,x_i) \ge
L(y_i,z_0) + L(z_0,x_i)$, $i = 1,2$, then either $x_1 = x_2$ and $y_1 =
y_2$ or $(x_1 - x_2)(y_1 - y_2) < 0$.

Suppose a sequence $\{x_n\}$, $n \in \zset$, has the following property:
for all $N_1$, $N_2 \in \zset$, $N_1 < N_2$, and any set $\{y_j\}$, $j =
N_1, N_1 + 1, \ldots, N_2$, such that $y_{N_1} = x_{N_1}$, $y_{N_2} =
x_{N_2}$,
$$
   \sum_{j = N_1}^{N_2 - 1} L(x_j,x_{j+1})
   \le \sum_{j=N_1}^{N_2 - 1} L(y_j,y_{j+1});
$$
then this sequence is called \emph{an $L$-minimal configuration}.

\begin{prop}
Suppose the function $L(y,x)$ satisfies the above conditions and $\{x_n\}$,
$n \in \zset$, is an $L$-minimal configuration; then there exists
\begin{equation}
   \lim_{n \to \pm\infty} {x_n - x_0\over n} = \omega(\{x_n\}) \in \rset.
\label{defomega}
\end{equation}
If $\omega = \omega(\{x_n\})$ is irrational, then there exists a function
$\phi_\omega(t)$:~$\rset \to \rset$ with the following properties:
(i) it is continuous on the right and $\phi_\omega(t_1) < \phi_\omega(t_2)$
iff $t_1 < t_2$;
(ii) $\phi_\omega(t + 1) = \phi_\omega(t) + 1$ for all $t \in \rset$;
(iii) if $x_0 = \phi_\omega(t_0 \pm 0)$ then the sequence $\{x_n\}$ defined
by $x_n = \phi_\omega(t_0 + n\omega \pm 0)$ is $L$-minimal;
(iv) for any neighborhood $V \in \mathcal{T}$ of the closure of the image
$\phi_\omega(\rset)$ and any $L$-minimal sequence $\{x_n\}$, $x_n \in
V$ as soon as $|n|$ is large enough.
\label{p:MF}
\end{prop}
For the proof see, e.g.,~\cite[sections~9,~11, and~12]{MF}. The number
$\omega(\{x_k\})$ is called a \emph{rotation number} of the $L$-minimal
configuration $\{x_k\}$.

Notice that $L_a(y,x)$ defined in the previous section satisfies the
conditions for $L(y,x)$.
\begin{prop}
Suppose $a \in \rset$, $s^a(x)$ is a continuous periodic function
satisfying~(\ref{functeq}), $Y^a(x)$ is the corresponding many-valued map,
$M^a$ is its invariant set, and $x \in M^a$. Then the sequence $x_k =
(Y^a)^{-k}(x)$, $k \in \zset$, is an $L$-minimal configuration. If $y_k =
(Y^a)^{-k}(y)$, $k \in \zset$, for some $y \in M^a$, then $\omega(\{x_k\})
= \omega(\{y_k\})$.
\label{p:rotnum}
\end{prop}
\textsc{Proof.} From~(\ref{ncc}) it follows that $s^a(x_{n+1}) = s^a(x_n) +
L_a(x_n,x_{n+1})$. Hence,
\begin{equation}
   s^a(x_{N_2}) - s^a(x_{N_1}) = \sum_{k = N_1}^{N_2 - 1} L_a(x_k,x_{k+1}).
\label{temp5}
\end{equation}
On the other hand,
$$
   s^a(x_{N_2}) = \min_{(y_{N_1},y_{N_1 + 1},\ldots,y_{N_2- 1}) \in
   \rset^{N_2 - N_1}}\,
   \left(s^a(y_{N_1}) + \sum_{k = N_1}^{N_2 - 1} L_a(y_k,y_{k+1})\right),
$$
where $y_{N_2} = x_{N_2}$. Thus for any set $\{y_j\}$, $j = N_1, N_1 + 1,
\ldots, N_2$, such that $y_{N_1} = x_{N_1}$, $y_{N_2} = x_{N_2}$, we have
\begin{equation}
   s^a(x_{N_2} - s^a(x_{N_1}) \le \sum_{k = N_1}^{N_2 - 1} L_a(y_k,y_{k+1}).
\label{temp6}
\end{equation}
Combining~(\ref{temp5}) with~(\ref{temp6}), we see that $\{x_n\}$ is
$L$-minimal.

Now let $y_k = (Y^a)^{-k}(y)$ for some $y \in M^a$. Take $p \in \zset$ such
that $x \le y + p \le x + 1$. By lemma~\ref{l:Yper}, it follows that
$(Y^a)^{-k}(x) \le (Y^a)^{-k}(y + p) \le (Y^a)^{-k}(x + 1)$ for all $k \in
\zset$. Using~(\ref{defomega}), we obtain $\omega(\{x_k\}) =
\omega(\{y_k\})$. \hfill$\square$

\section{Proof of theorem~II}

Let $a \in \rset$, $s^a(x)$ be a continuous periodic function
satisfying~(\ref{functeq}), and $M^a$ be the invariant set of the
corresponding map $Y^a(x)$. A subset $M \subset M^a$ is called \emph{a
minimal set} of the (single-valued) map $Y^a|_{M^a}$ if $M$ is closed,
$Y^a(M) = M$, and $M$ contains no other nonempty subset with the same
properties.  Let $x \in M^a$; the sequence $\{(Y^a)^k(x)\}$, $k =
1,2,\ldots$, is called \emph{the orbit} of the point $x$. It can be proved
that a set $M \subset M^a$ is minimal if and only if the orbit of any point
$x \in M$ is dense in $M$ in topology $\mathcal{T}$ defined in the proof of
proposition~\ref{p:Ma}.

\begin{lm}
Suppose $s^a_1(x)$ and $s^a_2(x)$ are continuous periodic functions
satisfying~(\ref{functeq}) and $Y^a(x)$ is the many-valued map
corresponding to $s^a_1(x)$; then for all $x$, $y \in \rset$ such that $y
\in Y^a(x)$
\begin{equation}
    s^a_1(x) - s^a_2(x) \ge s^a_1(y) - s^a_2(y).
\label{lastlm}
\end{equation}
\label{l:last}
\end{lm}
\textsc{Proof.} It follows from~(\ref{ncc}) that $s^a_1(x) = s^a_1(y) +
L_a(y,x)$. Using~(\ref{defsn}), we obtain
$$
\begin{array}{rl}
   s^a_1(x) - s^a_2(x)
   &\ds\!\!\!\!= s^a_1(y) + L_a(y,x) - s^a_2(x) = \\
   &\ds\!\!\!\!= s^a_2(y) + L_a(y,x) - s^a_2(x) + s^a_1(y) - s^a_2(y) \ge \\
   &\ds\!\!\!\!\ge s^a_1(y) - s^a_2(y).
\end{array}
$$
This completes the proof. \hfill$\square$

Let $s^a_1(x)$ and $s^a_2(x)$ be functions satisfying the assumptions of
lemma~\ref{l:last}, $Y^a_1(x)$ and $Y^a_2(x)$ be the corresponding
many-valued maps, and $M^a_1$ and $M^a_2$ be their invariant sets. By
proposition~\ref{p:rotnum}, it follows that the corresponding rotation
numbers $\omega_1$ and $\omega_2$ are determined uniquely. The set $M^a_1$
is compact in topology $\mathcal{T}$; using~(\ref{Yper}), we see that
$Y^a_1$ is a homeomorphism of $M^a_1$ in this topology. By the well-known
theorem of Birkhoff~\cite{B}, it follows that $M^a_1$ contains at least one
minimal set of the map $Y^a_1$; denote it by $M_1$.

Clearly, $s^a_1(x)$ and $s^a_2(x)$ are continuous in topology $\mathcal{T}$.
Let $x_0 \in M_1$ be a point where $s^a_1(x) - s^a_2(x)$ attains its
minimal value in $M_1$ and $\{x_k\}$, $k = 1,2,\ldots$, be its orbit. By
the above lemma,
$$
   s^a_1(x_0) - s^a_2(x_0) \le s^a_1(x_k) - s^a_2(x_k)
   \le s^a_1(x_0) - s^a_2(x_0).
$$
Since the closure of $\{x_k\}$ is the set $M_1$, we see that on $M_1$
$s^a_1(x) - s^a_2(x)$ is constant. Thus it follows from~(\ref{ncc}) that on
the set $M_1$ the map $Y^a_1(x)$ coincides with $Y^a_2(x)$. In particular,
$\omega_1 = \omega_2 = \omega$ or equivalently, $\omega$ depends only on
$a$. We see also that if $M$ is a minimal set of the map corresponding to
some $s^a(x)$ satisfying~(\ref{functeq}), then it is a minimal set of any
other continuous periodic function satisfying~(\ref{functeq}) with the same
$a$.

Now suppose $\omega = \omega(a)$ is irrational. Let us show that for any
continuous periodic function $s^a(x)$ satisfying~(\ref{functeq}) the
invariant set $M^a$ is minimal and hence uniquely determined. Indeed,
consider some $s^a(x)$; let $x \in M^a$ and $\{x_k\}$, $k = 1,2,\ldots$, be
its orbit. It follows from propositions~\ref{p:rotnum} and~\ref{p:MF} that
there exists $t_0 \in \rset$ such that $x_k = \phi_\omega(t_0 + k\omega \pm
0)$; in particular, this means that $M^a$ is the closure of the set
$\phi_\omega(\rset)$. Since $\omega$ is irrational, for any $t \in \rset$,
$\varepsilon > 0$ there exist $p_+$, $p_-$, $q_+$, $q_- \in \zset$, $q_+ >
0$, $q_- > 0$, such that $p_- - q_-\omega \in [t - \varepsilon, t]$, $p_+ -
q_+\omega \in [t, t + \varepsilon]$. Using strict monotonicity of
$\phi_\omega$, we see that the orbit of $x$ is dense in $M^a$ in topology
$\mathcal{T}$. This proves that $M^a$ is minimal.

Finally let us prove that if $\omega = \omega(a)$ is irrational, then
$s^a_1(x) - s^a_2(x)$ is constant for all $x \in \rset$. Without loss of
generality it can be assumed that $s^a_1(x) - s^a_2(x) = 0$ on $M^a$. Let
$x_0 \in \rset$ be a point where $s^a_1(x) - s^a_2(x)$ attains its minimum
in $\rset$ and $\{x_k\}$, $k = 1,2,\ldots$, be its orbit. Arguing as above,
we see that $s^a_1(x_k) - s^a_2(x_k) = s^a_1(x_0) - s^a_2(x_0)$ for all $k
\ge 1$.  On the other hand, it follows from proposition~\ref{p:Ma} that for
any neighborhood $V \in \mathcal{T}$ of the set $M^a$ there exists $N > 0$
such that $x_k \in V$ for all $k > N$. Using continuity of the functions
$s^a_1(x)$ and $s^a_2(x)$, we see that $s^a_1(x_0) - s^a_2(x_0) = 0$. By
definition of $x_0$, it follows that $s^a_1(x) \ge s^a_2(x)$ for all $x \in
\rset$.  Likewise, $s^a_1(x) \le s^a_2(x)$; thus $s^a_1(x) = s^a_2(x)$.

We have proved that if $\omega(a)$ is irrational, then the functional
equation~(\ref{functeq}) has a unique solution $s^a(x)$ in the class of
continuous periodic functions up to an additive constant. Thus the
solution~(\ref{FLO}) of the form~(\ref{defSa}) to the Cauchy problem for
equation~(\ref{HJE}) with the initial data $S_0(x) = ax + s^a(x)$ is
determined uniquely up to an additive constant.  This completes the proof
of theorem~II.

\end{document}